\documentclass[12pt,a4paper]{article}
\usepackage{amsmath,amssymb,theorem,cite,epsfig,url,psfrag,eepic,mathtools}
\usepackage{geometry}
\usepackage{amsmath}
\usepackage{feynmp}
\usepackage{wrapfig}
\usepackage{epsfig}
\usepackage{amssymb}
\usepackage{graphicx}
\usepackage{appendix}
\usepackage{slashed}
\usepackage{indentfirst}
\usepackage[T1]{fontenc}
\usepackage{color}
\usepackage{comment}
\usepackage[centertableaux,smalltableaux]{ytableau}
\topmargin by -\headsep \textheight 9.6in \oddsidemargin -25pt
\evensidemargin \oddsidemargin \marginparwidth 0.5in \textwidth
7.3in
\def\Maketitle{{\def\newpage{}\maketitle}}
\makeatletter
\def\Appendix{\appendix
  \def\@seccntformat##1{Appendix~\csname the##1\endcsname.~~}}
\makeatother
\makeatletter \@addtoreset{equation}{section}

\makeatother
\theorembodyfont{\sffamily}

\def\XXint#1#2#3{{\setbox0=\hbox{$#1{#2#3}{\int}$}
\vcenter{\hbox{$#2#3$}}\kern-.5\wd0}}

\def\�th{{\textrm{\,�th}}}

\newcommand{\ignore}[1]{}
\begin{document}

\title{\textbf{Coset conformal field theory \\ and
instanton counting on $\mathbb{C}^{2}/\mathbb{Z}_{p}$
}\vspace*{1cm}}
\author{M.~N.~Alfimov$^{1,2,3,4}$, A. A. Belavin$^{4,5,6}$,   G.~M.~ Tarnopolsky$^{6,7}$\vspace*{10pt}\\[\medskipamount]
$^1$~\parbox[t]{0.88\textwidth}{\small\it\raggedright
LPT, Ecole Normale Superieure, 24, rue Lhomond 75005 Paris, France}\\[\medskipamount]
$^2$~\parbox[t]{0.88\textwidth}{\small\it\raggedright
Institut de Physique Theorique, Orme des Merisiers, CEA Saclay 91191 Gif-sur-Yvette Cedex, France}\\[\medskipamount]
$^3$~\parbox[t]{0.88\textwidth}{\small\it\raggedright
P.N. Lebedev Physical Institute, Leninskiy pr. 53, 119991 Moscow, Russia}\\[\medskipamount]
$^4$~\parbox[t]{0.88\textwidth}{\small\it\raggedright
Moscow Institute of Physics and Technology, Institutskiy per. 9, 141700 Dolgoprudny, Russia}\\[\medskipamount]
$^5$~\parbox[t]{0.88\textwidth}{\small\it\raggedright
Institute for Information Transmission Problems, Bolshoi Karetnyi per. 19, 127994 Moscow, Russia}\\[\medskipamount]
$^6$~\parbox[t]{0.88\textwidth}{\small\it\raggedright Landau
Landau Institute for Theoretical Physics, Akademika Semenova pr. 1A, 142432 Chernogolovka, Russia}
\\[\medskipamount]
$^7$~\parbox[t]{0.88\textwidth}{\small\it\raggedright Department of Physics, Princeton University, Jadwin Hall, Washington Road, Princeton, NJ, 08540, U.S.A.}\vspace*{3pt}\\[\medskipamount]}

\date{}

\rightline{\texttt{\today}}

\rightline{LPT-ENS-13/15}

\Maketitle
\begin{abstract}\vspace*{10pt}

We study conformal field theory with the symmetry  algebra
$\mathcal{A}(2,p)=\widehat{\mathfrak{gl}}(n)_{2}/\widehat{\mathfrak{gl}}(n-p)_2$.
In order to support the conjecture that this algebra acts on the
moduli space of instantons on $\mathbb{C}^{2}/\mathbb{Z}_{p}$, we
calculate the characters of its representations and check their
coincidence with the generating functions of the fixed points of the
moduli space of instantons.

We show that the algebra $\mathcal{A}(2,p)$ can be realized in two
ways. The first realization is connected with the cross-product of
$p$ Virasoro and $p$ Heisenberg algebras: $\mathcal{H}^{p}\times
\textrm{Vir}^{p}$. The second realization is connected with:
 $\mathcal{H}^{p}\times \widehat{\mathfrak{sl}}(p)_2\times (\widehat{\mathfrak{sl}}(2)_p \times
 \widehat{\mathfrak{sl}}(2)_{n-p}/\widehat{\mathfrak{sl}}(2)_n )$. The equivalence of these
two realizations provides the non-trivial identity for the
characters of $\mathcal{A}(2,p)$.

The moduli space of instantons on $\mathbb{C}^{2}/\mathbb{Z}_{p}$
admits two different compactifications. This leads to two different
bases for the representations of $\mathcal{A}(2,p)$. We use this fact to
explain the existence of two forms of the instanton pure partition
functions.

\end{abstract}

\tableofcontents

\section{Introduction}

In recent years the remarkable relation between two dimensional
conformal field theories and four dimensional $\mathcal{N}=2$
supersymmetric Yang-Mills theories  experienced a considerable
development. The  original so-called AGT relation proposed in
\cite{Alday:2009aq} states the equality between the correlation
functions in the Liouville field theory and the partition function
of the $\mathcal{N}=2$ supersymmetric Yang-Mills theory with the
$SU(2)$ gauge group (for the generalizations of the AGT correspondence for the other gauge groups and conformal field theories see \cite{Wyllard:2009hg,Mironov:2009by,Alday:2010vg,Belavin:2011pp,Tan:2013tq,Belavin:2011tb,Ito:2011mw,Belavin:2012aa,Wyllard:2011mn,Alfimov:2011ju,Belavin:2012uf}). The partition function of the $\mathcal{N}=2$
supersymmetric gauge theory can be calculated as the integral over
the moduli space of instantons $\mathcal{M}$. With  the proper
regularization,  this integral was computed explicitly
\cite{Nekrasov:2003rj}. This was achieved using the localization
theorem, which shows that the integral is fully determined by the
fixed points of some abelian group (torus) acting on the moduli
space $\mathcal{M}$  \cite{Flume:2002az}.

In the present work we consider the $U(r)$ instantons  on
$\mathbb{C}^2/\mathbb{Z}_p$~--- the solutions of the self-duality
equation, with the additional condition for the gauge field:
\begin{equation}
  A_{\mu}(z_1,z_2) = A_{\mu}(\omega z_1, \omega^{-1} z_2), \quad  \omega^p=1.
\end{equation}
The nontrivial fact about the moduli space of instantons
$\mathcal{M}$  is that one can construct the action of some symmetry
algebra $\mathcal{A}$ on the equivariant cohomologies of the moduli
space $\mathcal{M}$. The first examples of such action were given by
Nakajima in \cite{1995alg.geom..7012N,1999math.....12158N} for the
cases of Heisenberg and Kac-Moody algebras. In \cite{Atiyah:1984px}
it was shown that the basis in the space of the equivariant
cohomologies can be labelled by the fixed points of the torus acting
on the moduli space. Therefore it is natural to assume the existence of the special basis of the geometrical origin in the representation of $\mathcal{A}$,
elements of which being in one to one correspondence with the torus
fixed points on the moduli space. This basis has a number of
remarkable properties, which are listed in \cite{Belavin:2011bb}.

In the work \cite{Belavin:2011pp} it was suggested that the
instanton moduli  space of the $\mathcal{N}=2$ supersymmetric $U(r)$
gauge theory on $\mathbb{C}^2/\mathbb{Z}_p$ is connected to the
algebra $\mathcal{A}(r,p)$, which is realized by the coset
\begin{equation}\label{A_definition}
  \mathcal{A}(r,p) \mathop{=}^{\textrm{def}}
  \frac{\widehat{\mathfrak{gl}}(n)_r}{\widehat{\mathfrak{gl}}(n-p)_r},
\end{equation}
where $n$ is related to the equivariant parameters (to learn more
details about this correspondence see \cite{Belavin:2011bb}). In
other words there exists the special basis in the representation of $\mathcal{A}(r,p)$,
whose elements are in one to one correspondence with the fixed points
of the torus acting on the instanton moduli space
$\mathcal{M}$. On the other hand these fixed points can be
enumerated by the $r$-tuples of the Young diagrams painted in $p$
colors. Therefore we can associate the specific $r$-tuple of colored Young diagrams to each element of this geometrical basis. Such bases were explicitly constructed for $r=2$ and $p=1$ in
\cite{Alba:2010qc} and for $r=2$ and $p=2$ in \cite{Belavin:2011bb}
and for $r=1,2$ and $p=2$ in \cite{Belavin:2012eg}.

The present work can be considered as the continuation of the line
of studies started in \cite{Belavin:2011bb,Alfimov:2011ju,Ito:2011mw,Belavin:2012eg,Belavin:2012aa}. The main
goal of this work is to find nontrivial evidence in support of the
conjectured correspondence between the algebra $\mathcal{A}(2,p)$
and the moduli space of $U(2)$ instantons on
$\mathbb{C}^2/\mathbb{Z}_p$, which we denote by $\bigsqcup_{N}
\mathcal{M}(2,N)^{\mathbb{Z}_p}$. Namely, we check the
correspondence between the fixed points on the moduli space of
$U(2)$ instantons on $\mathbb{C}^2/\mathbb{Z}_p$  and the vectors in
the representation of the algebra $\mathcal{A}(2,p)$, by comparing the generating functions of the fixed points and the characters of
the representations.

 Using the level-rank duality the algebra $\mathcal{A}(2,p)$ can be represented in two ways
\begin{equation} \label{algebra_corr}
\begin{picture}(200,90)(150,0)
\put(10,28){\mbox{\normalsize{$\bigsqcup_N\mathcal{M}(2,N)^{\mathbb{Z}_p}$}}}
\put(110,33){\vector(1,0){15}} \put(110,33){\vector(-1,0){15}}
\put(140,28){\mbox{\normalsize{$\mathcal{A}(2,p)$}}}
\put(190,35){\vector(1,1){20}} \put(190,28){\vector(1,-1){20}}
\put(220,55){\mbox{\small{$(\mathcal{H} \times \textrm{Vir}^{(1)})
\times \ldots \times (\mathcal{H} \times \textrm{Vir}^{(p)})$}}}
\put(220,5){\mbox{\large{$ \frac{\widehat{\mathfrak{sl}}(2)_p \times
    \widehat{\mathfrak{sl}}(2)_{n-p}}{\widehat{\mathfrak{sl}}(2)_n}$}\footnotesize{$ \times
\mathcal{M}(3/4) \times \ldots \times
\mathcal{M}(p+1/p+2)\times\mathcal{H}^p$}}}
    \put(310,33){\vector(0,1){12}}
    \put(310,33){\vector(0,-1){12}}
     \put(108,35){$1$}
     \put(192,48){$2$}
      \put(192,12){$3$}
      \put(299,29){$4$}
\end{picture}
\end{equation}

\vspace{0.5cm} \noindent where $\textrm{Vir}^{(\sigma)}$,
$\sigma=1,...,p$  are the Virasoro algebras with the special central
charges $c_{\sigma}$, $\mathcal{H}$ is the Heisenberg algebra,
$\mathcal{M}(m/m+1)$ is the Minimal model\footnote{Here we have
already used another level-rank duality, which gives
$\widehat{\mathfrak{sl}}(p)_2 = \mathcal{M}(3/4) \times \ldots
\times \mathcal{M}(p+1/p+2)$.} and $\widehat{\mathfrak{sl}}(2)_p
\times
\widehat{\mathfrak{sl}}(2)_{n-p}/\widehat{\mathfrak{sl}}(2)_n$ is
the coset algebra. In this paper we study the connections
(\ref{algebra_corr}) in detail.

In the Section \ref{fixed_points_counting} we find the generating
function for the  fixed points of the moduli space. In the Section
\ref{first_realization} we study the first realization of
$\mathcal{A}(2,p)$ depicted by the arrow $2$ on (\ref{algebra_corr})
and elaborate on the characters of the representation of
$\mathcal{A}(2,p)$ in this realization and compare them with the
generating functions of the fixed points of the moduli space. In the
Section \ref{second_realization} we study the second realization of
$\mathcal{A}(2,p)$ represented by the arrow $3$ on
(\ref{algebra_corr}). Also we find the coincidence of the characters
of the first and second realizations of $\mathcal{A}(2,p)$
(arrow $4$ on (\ref{algebra_corr})). In the Section
\ref{partition_functions}  we find the equalities between the
instanton partition functions of the $\mathcal{N}=2$ supersymmetric
$U(2)$ pure gauge theory on $\mathbb{C}^2/\mathbb{Z}_p$ calculated
for the different compactifications of the moduli space.

\section{Counting of the torus fixed points on the moduli space of instantons} \label{fixed_points_counting}

In this Section we concentrate on the counting of the fixed points
of the  moduli space. Let us consider the compactification of the
moduli space, where the action of the $\mathbb{Z}_p$ is lifted to
the moduli space of instantons on $\mathbb{C}^2$. In this case it is
convenient to label the fixed points by the $r$-tuples of Young
diagrams with $p$ colors. We are going to introduce the generating
functions of such Young diagrams and study the properties of such
generating functions.

\subsection{Fixed points on the moduli space of $U(2)$ instantons on $\mathbb{C}^2/\mathbb{Z}_p$}

Here we confine ourselves to the case of $r=2$, which means that we
consider the $U(2)$ instantons on $\mathbb{C}^2/\mathbb{Z}_p$, whose
 moduli space is $\bigsqcup_{N}
\mathcal{M}(2,N)^{\mathbb{Z}_p}$. It is convenient to numerate the torus fixed points in this
case by the pairs of Young diagrams
colored in $p$ colors. The coloring goes as follows. We ascribe the
color $r$ from $0$ to $p-1$ to the corner cell and the colors $r+i-j
\mod p$ to the cell with the coordinates $(i,j)$. For example, the
diagram with $r=3$ and $p=4$

\begin{equation*}
\begin{ytableau}
3 & 0 & 1 & 2 & 3 & 0 \\
2 & 3 & 0 & 1 \\
1 & 2 & 3 & 0 \\
0 & 1 \\
3
\end{ytableau}
\end{equation*}

In this section we count these fixed points. Because the number of
the pairs  of Young diagrams colored in $p$ colors is infinite, we
need to introduce the grading of these diagrams. One possible way to
do that is to count the pairs of Young diagrams with the fixed
common size, fixed colors of the corner cells $r_1$, $r_2$ and
differences $k_m$ between the number of cells with the color $m>0$
and the number of cells with the color $0$ in both diagrams. We will
call the set of numbers $r_1$, $r_2$ and $k_{1},\ldots,k_{p-1}$
($k_0=0$ by definition) as the coloring of the Young diagrams. Thus,
we determine the generating function of the pair of Young diagrams
as follows
\begin{equation}
\chi_{r_1,r_2}(k_1,\ldots,k_{p-1}|q)=\sum_{(Y_1,Y_2) \in \triangledown}q^{\frac{|Y_1|+|Y_2|}{p}},
\end{equation}
where
\begin{equation}
\triangledown=\{(Y_{1}, Y_{2})|\;
\begin{ytableau}
r_1 & & & \\
& & \\
\
\end{ytableau}
,
\begin{ytableau}
r_2 & & \\
& \\
\
\end{ytableau},
\sharp(
\begin{ytableau}
m
\end{ytableau})-\sharp(
\begin{ytableau}
0
\end{ytableau})=k_{m}\}.
\end{equation}
and $|Y|$ is the number of the boxes in the Young diagram $Y$.

It should be noted here that the coloring parameters of the Young
diagrams are connected with the topological characteristics of the
instantons corresponding to the fixed points of torus action on the
moduli space. Let $c_1(E)$ be the first Chern class of the gauge
bundle and $c_1(T_r)$ be the first Chern class of the vector bundle
on the ALE space, then
\begin{equation}
c_1(E)=\sum_{r=1}^{p-1}c_{(r)}c_1(T_r).
\end{equation}
Denoting the number of Young diagrams with the color of corner cell $r$ by $n_r$ we get
\begin{equation}\label{Chern_class}
c_{(r)}=n_r+k_{r+1}-2k_r+k_{r-1}, \; r=1,\ldots,p-1.
\end{equation}
Utilizing (\ref{Chern_class}) one can pass from the description  of the
generating functions of the Young diagrams in terms of colorings to
the equivalent description in terms of Chern classes.

The generating function of the pair of colored Young diagrams  can
be constructed from the generating function of one colored Young
diagram, which can be extracted from \cite{2005math.....10455F}. Let
$r$ be the color of the corner cell, then for $r=0,...,p-1$  the
generating function of one colored Young diagram is defined as
\begin{align}
\chi_{r}(k_{1},...,k_{p-1}|q) \mathop{=}^{\textrm{def}} \sum_{Y \in \lozenge} q^{\frac{|Y|}{p}},
\end{align}
where $\lozenge$ is  the set of Young diagrams with the particular
coloring
\begin{align}
\lozenge= \{Y|\;
\begin{ytableau}
r & & & \\
& & \\
\
\end{ytableau}
,\sharp(\begin{ytableau} m
\end{ytableau})-\sharp(\begin{ytableau}
0
\end{ytableau})=k_{m}\},
\end{align}

\noindent where the box with coordinates $(i,j)$ has the color $r+i-j \, \textrm{mod} \, p$ and
$
\sharp(\begin{ytableau}
m
\end{ytableau}),\sharp(\begin{ytableau}
0
\end{ytableau})$~--
the numbers of the boxes with $m$ and $0$ colors respectively. For
example we have
\begin{align}
\chi_{2}(0,1|q) &= q^{\frac{1}{3}}\cdot \sharp\{\ytableaushort{2}\} + q^{\frac{4}{3}} \cdot \sharp \{\;\ytableaushort{20,12},\ytableaushort{2012},\ytableaushort{2,1,0,2}\;\}+O(q^{\frac{7}{3}})=\notag\\
&=q^{\frac{1}{3}}+3 q^{\frac{4}{3}}+O(q^{\frac{7}{3}}).
\end{align}
Introducing a convenient notation for the character of the highest
weight representation of the Heisenberg algebra\footnote{ The
Heisenberg algebra consists of the elements $a_{k}$ without $a_0$
and has the commutation relations $[a_{n},a_{m}]=n\delta_{n+m,0}$.
The highest weight representation of this algebra (Fock module) has
the vacuum state $|0\rangle$: $a_{n}|0\rangle =0$ for $n>0$, and is
spanned by the vectors $a_{-n_{1}}...a_{-n_{k}}|0\rangle$,
$n_{1}\geqslant n_{2} \geqslant...\geqslant n_{k}$.} $\mathcal{H}$
\begin{align}
\chi_{B}(q)= \prod_{n=1}^{\infty}\frac{1}{1-q^{n}}, \label{bosc}
\end{align}
one can get for the generating function of  one colored Young
diagram (here we imply $k_{0}=k_{p}=0$)
\begin{align}
\chi_{r}(k_{1},...,k_{p-1}|q)=q^{\sum_{i=1}^{p-1}(k_{i}^{2}
+\frac{k_{i}}{p}-k_{i}k_{i+1})-k_{r}}\cdot (\chi_{B}(q))^{p}.
\label{colchar1}
\end{align}

It is straightforward to obtain the generating function of the pair
of Young diagrams. The generating function of the pair of Young diagrams with the coloring $r_1,r_2$ and
$k_1,\ldots,k_{p-1}$ is equal to
\begin{align}
\chi_{r_{1},r_{2}}(k_{1},...,k_{p-1}|q)=
\sum_{\substack{m_{i}+n_{i}=k_{i} \\
i=1,...,p-1 }}   \chi_{r_{1}}(m_{1},...,m_{p-1}|q)
\chi_{r_{2}}(n_{1},...,n_{p-1}|q).
\end{align}
Then using the formula (\ref{colchar1}) one obtains
\begin{align}
&\chi_{r_{1},r_{2}}(k_{1},...,k_{p-1}|q)= \notag\\
&\qquad=(\chi_{B}(q))^{2p}\sum_{m_{1},...,m_{p-1} \in \mathbb{Z}}
q^{\frac{1}{2}\sum_{i=1}^{p-1}((2m_{i}-k_{i})^{2}-(2m_{i}-k_{i})(2m_{i+1}-k_{i+1})
+k_{i}^{2}-k_{i}k_{i+1}+\frac{2k_{i}}{p})-m_{r_{1}}+m_{r_{2}}-k_{r_{2}}}.
\label{charc}
\end{align}
And therefore we can obviously get
\begin{align}
&\sum_{k_{1},...,k_{p-1}=0}^{1}q^{-\frac{1}{2}\sum_{i=1}^{p-1}(k_{i}^{2}-k_{i}k_{i+1}+\frac{2k_{i}}{p})+\frac{1}{2}(k_{r_{1}}+k_{r_{2}})}
\chi_{r_{1},r_{2}}(k_{1},...,k_{p-1}|q)= \notag\\
&\qquad\qquad\qquad\qquad=(\chi_{B}(q))^{2p}\sum_{n_1, \ldots,
n_{p-1} \in \mathbb{Z} \atop
    n_0=n_p=0} q^{\frac{1}{2}\sum_{\sigma=1}^{p-1}
    (n_{\sigma}^2-n_{\sigma}n_{\sigma+1})+\frac{1}{2}(n_{r_{2}}-n_{r_{1}})}.
\label{charcn}
\end{align}

\subsection{Counting of the non equivalent generating functions of the colored Young diagrams} \label{counting_color_char}

We call two generating functions equivalent if they differ only by the
multiplication of $q$ to some power (by this definition all the
generating functions (\ref{colchar1}) are equivalent). Below we show
that the whole infinite set of the generating functions
(\ref{charc}) can be divided into the finite number of classes of
equivalence. From (\ref{charc}) we conclude that the generating
functions  have the following symmetries:
\begin{itemize}
\item The invariance under the transformation $k_m \to k_m+2$:
\begin{equation} \label{symmetry1}
\chi_{r_1,r_2}(k_1,...,k_m+2,...,k_{p-1}|q)
=q^{2k_m-k_{m+1}-k_{m-1}+\frac{2}{p}+\delta_{m,r_1}+\delta_{m,r_2}}
\chi_{r_1,r_2}(k_1,...,k_{p-1}|q),
\end{equation}
where $\delta_{m,n}$ is the Kronecker delta.
\item The invariance under the permutation $r_1 \leftrightarrow r_2$:
\begin{equation} \label{symmetry2}
\chi_{r_1,r_2}(k_1,...,k_{p-1}|q)=\chi_{r_2,r_1}(k_1,...,k_{p-1}|q).
\end{equation}
\item The invariance under the change
$r_1,r_2 \rightarrow r_1+1,r_2+1$:
\begin{equation} \label{symmetry3}
\chi_{r_1+1,r_2+1}(k_1,...,k_{p-1}|q)
=q^{k_{r_1}-k_{r_1+1}-\frac{r_2-r_1}{p}}\chi_{r_1,r_2}(k_1,...,k_{r_1+1}+1,...,k_{r_2}+1,...,k_{p-1}|q),
\end{equation}
where we assume that $r_{1} \leqslant r_{2}$.

\end{itemize}

Applying the symmetries (\ref{symmetry1} - \ref{symmetry3}) we
conclude that  an arbitrary generating function  with the coloring
$r_1,r_2$ and $k_1,...,k_{p-1}$ is equivalent to one of the generating functions
with $r_1=0$ and all $k_m$'s equal to 0 or 1:
\begin{equation} \label{app_sym_gen_func}
\chi_{0,s}(k_1,...,k_{p-1}|q),
\end{equation}
where $s=0,1,...,p-1$. Therefore we can confine our  consideration
to the generating functions of the form (\ref{app_sym_gen_func}).

It can be shown (see Appendix \ref{app_A}) that for each
$s=0,\ldots,p-1$ the generating functions (\ref{app_sym_gen_func})
are divided into $[s/2]+[(p-s)/2]+1$ classes of
equivalence\footnote{The symbol [...] means integer part of a
number, for example [3/2]=1. }. The first class of equivalence
contains the generating functions equivalent to
\begin{equation} \label{rep_class_1}
\chi_{0,s}(0,...,0|q),
\end{equation}
and its cardinality is  $\binom{p}{s} = \frac{p!}{s!(p-s)!}$.  For
each of the next $[s/2]$ classes of equivalence it is convenient to
choose the representative
\begin{equation}\label{rep_class_2}
\chi_{0,s}(0,...,\stackrel{s-2l+1}{0,1,0},1,0,...,1,0,\stackrel{s}{1,0,0},...,0|q),
\end{equation}
where $l$ takes integer values from $1$ to $[s/2]$, and from
$k_{s-2l+1}$ to $k_{s}$ there are alternating 1 and 0 and all other
$k_i=0$. The cardinality of the equivalence class with given $l$ is
$\binom{p}{s-2l}$. And for each of the last $[(p-s)/2]$ classes of
equivalence we choose the representative
\begin{equation}\label{rep_class_3}
\chi_{0,s}(0,...,\stackrel{s}{0,0,1},0,1,...,0,1,\stackrel{s+2n-1}{0,1,0},...,0|q),
\end{equation}
where $n$ takes the values from $1$ to $[(p-s)/2]$, and from $k_{s}$
to $k_{s+2n-1}$ there are alternating 0 and 1 and all other $k_i=0$.
One can easily check that the sum of the cardinalities of the
equivalence classes indeed equals to the number of generating
functions (\ref{app_sym_gen_func}) with given $s$
\begin{equation}
\binom{p}{s}+\sum_{l=1}^{[s/2]}\binom{p}{s-2l}+\sum_{n=1}^{[(p-s)/2]}\binom{p}{p-s-2n}=2^{p-1}.
\end{equation}
Now we see that arbitrary generating functions of the fixed points
on the moduli space is equivalent to one of the generating functions
(\ref{rep_class_1} - \ref{rep_class_3}) and we obtain
\begin{align}
\sum_{k_{1},...,k_{p-1}=0}^{1}&q^{-\frac{1}{2}\sum_{i=1}^{p-1}(k_{i}^{2}-k_{i}k_{i+1}+\frac{2k_{i}}{p})+\frac{k_{s}}{2}}
\chi_{0,s}(k_{1},...,k_{p-1}|q) =
\binom{p}{s} \chi_{0,s}(0,...,0|q)+ \notag \\
&\qquad +\sum_{n=1}^{\left[s/2 \right]} \binom{p}{s-2n}
q^{-\frac{n}{2}\left(1+\frac{2}{p}
\right)}\chi_{0,s}(0,...,0,\stackrel{s-2n+1}{0,1,0},
...,1,0,\stackrel{s}{1,0,0},...,0|q)+ \notag \\
&\qquad +\sum_{n=1}^{\left[(p-s)/2 \right]}\binom{p}{s+2n}
q^{-\frac{n}{2} \left(1+\frac{2}{p} \right)}
\chi_{0,s}(0,...,\stackrel{s}{0,0,1},0,1,...,\stackrel{s+2n-1}{0,1,0},0,
...,0|q). {\label{supident}}
\end{align}

\section{First realization of the algebra $\mathcal{A}(2,p)$} \label{first_realization}

In the present section we study the realization of the
$\mathcal{A}(2,p)$ as the product of $p$ models with Virasoro
symmetry. This realization is represented by the arrow 2 on the
figure (\ref{algebra_corr}). Let us start with the definition of
$\mathcal{A}(r,p)$
\begin{equation}
\mathcal{A}(r,p)=\frac{\widehat{\mathfrak{gl}}(n)_r}{\widehat{\mathfrak{gl}}(n-p)_r}. \label{Arp1}
\end{equation}
One should notice that the definition (\ref{Arp1})  makes sense only
for the positive integer values of the parameter $n$, but the usage
of the level-rank duality allows us to extend the definition of
$\mathcal{A}(r,p)$ to the case of arbitrary complex $n$. Formally
multiplying and dividing (\ref{Arp1}) by the algebras
$\widehat{\mathfrak{gl}}(n-\sigma+1)_r$ with $\sigma=1,....,p$ we
can write
\begin{equation}\label{A_definition1}
    \mathcal{A}(r,p)\supset \frac{\widehat{\mathfrak{gl}}(n-p+1)_r}{\widehat{\mathfrak{gl}}(n-p)_r}
    \times ... \times \frac{\widehat{\mathfrak{gl}}(n)_r}{\widehat{\mathfrak{gl}}(n-1)_r}.
\end{equation}
Multiplying and dividing each multiplier of (\ref{A_definition1}) by $\widehat{\mathfrak{gl}}(1)_r $, and using the level-rank duality\footnote{Here we used the following level-rank duality trick: $\frac{\widehat{\mathfrak{gl}}(k)_r}{\widehat{\mathfrak{gl}}(1)_r \times \widehat{\mathfrak{gl}}(k-1)_r} =\frac{\widehat{\mathfrak{sl}} (r)_{1} \times \widehat{\mathfrak{sl}}(r)_{k-1}}{\widehat{\mathfrak{sl}}(r)_{k}} $ and also the isomorphism $\widehat{\mathfrak{gl}}(1)_r  \cong \mathcal{H}$.} we get
\begin{equation}\label{A_definition2}
\mathcal{A}(r,p)\supset \left( \mathcal{H} \times \frac{\widehat{\mathfrak{sl}}(r)_{1} \times\widehat{\mathfrak{sl}}(r)_{n-p}}{\widehat{\mathfrak{sl}}(r)_{n-p+1}} \right) \times \ldots \times \left( \mathcal{H} \times \frac{\widehat{\mathfrak{sl}}(r)_{1} \times \widehat{\mathfrak{sl}}(r)_{n-1}}{\widehat{\mathfrak{sl}}(r)_{n}} \right).
\end{equation}
Note that the cosets $\widehat{\mathfrak{sl}}(r)_{1} \times
\widehat{\mathfrak{sl}}(r)_{n-\sigma}/\widehat{\mathfrak{sl}}(r)_{n-\sigma+1}$
for $\sigma=1,...,p$ are the $\textrm{W}_r$ theories with the
central charges
\begin{equation}\label{}
    c_{\sigma}=1+\frac{(r^2-1)(n-\sigma)}{n-\sigma+r}-\frac{(r^2-1)(n-\sigma+1)}{n-\sigma+r+1}.
\end{equation}

In the case of $r=2$ we have for (\ref{A_definition2}):
\begin{align}
    & \mathcal{A}(2,p)\supset \mathcal{H}^p \times \textrm{Vir}^{(1)}\times...\times\textrm{Vir}^{(p)},
    \label{Virasoro representation}
\end{align}
where $\mathcal{H}^{p}= \mathcal{H}\times...\times \mathcal{H}$ and
due to GKO correspondence \cite{Goddard:1984vk} the cosets
$\widehat{\mathfrak{sl}}(2)_{1}
\times\widehat{\mathfrak{sl}}(2)_{n-\sigma}/\widehat{\mathfrak{sl}}(2)_{n-\sigma+1}$
are the Virasoro algebras $\textrm{Vir}^{(\sigma)}$ with the
following  central charges (see Appendix \ref{app_cosets})
\begin{equation}\label{jVirasoro_parameters}
    c_{\sigma}=1+\frac{3(n-\sigma)}{n-\sigma+2}-\frac{3(n-\sigma+1)}{n-\sigma+3}=1+6(Q_{\sigma})^2,
\end{equation}
where $Q_{\sigma}=b_{\sigma}+b_{\sigma}^{-1}$ and $b_{\sigma}^2=-\frac{n-\sigma+3}{n-\sigma+2}$ is the parametrization, which will be useful in what follows. It is easy to check that the parameters $b_{\sigma}$ satisfy the following relations
\begin{equation}\label{b_relation}
b_{\sigma}^2+b_{\sigma+1}^{-2}=-2, \quad  \sigma=1,...,p-1.
\end{equation}
Thus, we have $p-1$ equations for $p$ variables. This means that we are able to express these variables in terms of the only one variable $b$
\begin{equation}\label{b_sigma_def}
b_{\sigma}^{2} = \frac{(\sigma-1)Q-pb}{\sigma Q - pb}, \quad
    \sigma=1,...,p.
\end{equation}

In the next subsection we enlarge the r.h.s of (\ref{Virasoro
representation}) up to the full $\mathcal{A}(2,p)$ algebra. In order
to perform this we add the set of $p-1$ holomorphic currents of spin
$1/2$ to the $p$ stress-energy tensors.

\subsection{The  $p$ models with Virasoro symmetry}

In the case when parameter $n$ in
$\mathcal{A}(2,p)=\widehat{\mathfrak{gl}}(n)_2 /
\widehat{\mathfrak{gl}}(n-p)_2$  is a positive integer, the cosets
$\widehat{\mathfrak{sl}}(2)_{1}
\times\widehat{\mathfrak{sl}}(2)_{n-\sigma}/\widehat{\mathfrak{sl}}(2)_{n-\sigma+1}$
in (\ref{A_definition2}) describe the Minimal models
$\mathcal{M}(n-\sigma+1/n-\sigma+2)$ and the arrow $4$ of the scheme
(\ref{algebra_corr}) exactly reproduces the correspondence between
the Minimal models, which was widely studied in
\cite{Crnkovic:1989ug}. In this section in the spirit of
\cite{Crnkovic:1989ug} we construct the first realization of the
algebra $\mathcal{A}(2,p)$.

We recall that  Virasoro algebra consists of the infinite number of
generators $L_n$, $n \in \mathbb{Z}$ satisfying the following
commutation relations
\begin{equation}
[L_n,L_m]=(n-m)L_{n+m}+\frac{c}{12}(n^3-n)\delta_{n+m,0},
\end{equation}
where $c$ is the central charge parametrized as $c=1+6Q^2$  with
$Q=b+b^{-1}$. We denote the highest weight state of this algebra by
$V_{\lambda}$, which is annihilated by $L_n$ with $n>0$ and has the
conformal dimension
\begin{equation}
\Delta(\lambda)=\frac{Q^2}{4}-\lambda^2.
\end{equation}
The highest weight states of the degenerate  representations of the
Virasoro algebra are denoted by $V_{m,n} \equiv V_{\lambda_{m,n}}$,
where
\begin{equation}
\lambda_{m,n}=\frac{mb^{-1}+nb}{2}
\end{equation}
and their dimension is
\begin{equation}
\Delta_{m,n}=\frac{Q^2}{4}-\lambda_{m,n}^2.
\end{equation}

In addition to $p$ stress-energy tensors $T^{(\sigma)}$ we form the
set of $p-1$ holomorphic currents
\begin{align}
J^{(\sigma)}(z)\mathop{=}^{\textrm{def}}V^{(\sigma)}_{1,2}(z) V^{(\sigma+1)}_{2,1}(z), \quad \sigma=1,...,p-1,
\end{align}
where $V^{(\sigma)}_{m,n}$ is the degenerate field primary with
respect to $T^{(\sigma)}$. Due to the relation (\ref{b_relation})
for $b_{\sigma}$'s the left conformal dimension of the current
$J^{(\sigma)}(z)$ is
\begin{equation}
\Delta_{J^{(\sigma)}}=\Delta_{1,2}^{(\sigma)}+\Delta_{2,1}^{(\sigma+1)}=\frac{1}{2},
\end{equation}
while the right conformal dimension is zero. It can be checked that
the currents $T^{(\sigma)}(z)$ and $J^{(\sigma)}(z)$ generate an
associative  chiral algebra \cite{Crnkovic:1989ug, Spodyneiko:2012}.
We call this algebra the first realization of the
$\mathcal{A}(2,p)$.

Now let us turn to the construction of the representations of the
$\mathcal{A}(2,p)$. The first requirement for the highest weight
state of the algebra is that it has to be primary with respect to
$p$ stress tensors $T^{(\sigma)}$. If
$V^{(\sigma)}_{\lambda_{\sigma}}$ is the primary state of the
$\sigma$-th stress tensor, then evidently the state
\begin{equation}
V^{(1)}_{\lambda_1} V^{(2)}_{\lambda_2} \ldots V^{(p)}_{\lambda_p}
\label{pst}
\end{equation}
is primary with respect to all stress tensors.

 Consider the OPE of the currents $J^{(\sigma)}(z)$ with the state
$V^{(1)}_{\lambda_{1}}...V^{(p)}_{\lambda_{p}}$. From the fusion rules
it follows
\begin{multline}\label{J_current_OPE1}
    J^{(\sigma)}(z)V^{(1)}_{\lambda_{1}}(0)... V^{(p)}_{\lambda_{p}}(0)=\sum_{m_{\sigma},m_{\sigma+1}=\pm 1}
    z^{m_{\sigma}\lambda_{\sigma} b_{\sigma}+m_{\sigma+1}\lambda_{\sigma+1} b_{\sigma+1}^{-1}} \times \\
    \times C^{(\sigma)}(m_{\sigma},m_{\sigma+1};\lambda_1,\ldots,\lambda_p) \left[ V^{(1)}_{\lambda_{1}} ... V^{(\sigma)}_{\lambda_{\sigma}
    +\frac{m_{\sigma} b_{\sigma}}{2}}
    V^{(\sigma+1)}_{\lambda_{\sigma+1}+\frac{m_{\sigma+1}}{2b_{\sigma+1}}}...V^{(p)}_{\lambda_{p}} \right],
\end{multline}
where
$C^{(\sigma)}(m_{\sigma},m_{\sigma+1};\lambda_1,\ldots,\lambda_p)$
are the structure constants. To reach locality we have to make the
projection\cite{Crnkovic:1989ug} and keep only two terms in the sum
(\ref{J_current_OPE1}), say, with $m_{\sigma}=m_{\sigma+1}= \pm 1$
and also impose condition $\lambda_{\sigma}
b_{\sigma}+\lambda_{\sigma+1} b_{\sigma+1}^{-1} \in \mathbb{Z}$ or
$\mathbb{Z}+1/2$. Now modes of $J^{(\sigma)}$, which act on the
states of the representation, are correspondingly half-integer or
integer
\begin{align*}
& J^{(\sigma)}(z)V^{(1)}_{\lambda_{1}}(0)... V^{(p)}_{\lambda_{p}}(0)=
\sum_{n \in \mathbb{Z}+\frac{1}{2}} z^{n-\frac{1}{2}} J^{(\sigma)}_{n}V^{(1)}_{\lambda_{1}}
(0)... V^{(p)}_{\lambda_{p}}(0), \;\;\; \textrm{if} \;\;\; \lambda_{\sigma}b_{\sigma}+\lambda_{\sigma+1}b_{\sigma+1}^{-1} \in \mathbb{Z}, \\
& J^{(\sigma)}(z)V^{(1)}_{\lambda_{1}}(0)...
V^{(p)}_{\lambda_{p}}(0) =\sum_{n \in \mathbb{Z}} z^{n-\frac{1}{2}}
J^{(\sigma)}_{n}V^{(1)}_{\lambda_{1}}(0)...
V^{(p)}_{\lambda_{p}}(0), \;\;\;
 \textrm{if} \;\;\; \lambda_{\sigma}b_{\sigma}+\lambda_{\sigma+1}b_{\sigma+1}^{-1} \in \mathbb{Z}+\frac{1}{2}.
\end{align*}
The second requirement is that the state (\ref{pst}) is primary for the
currents $J^{(\sigma)}$, i.e. it's  annihilated by the modes of all
currents $J^{(\sigma)}(z)$ with positive numbers
\begin{equation}
J^{(\sigma)}_n V^{(1)}_{\lambda_{1}}(0)...
V^{(p)}_{\lambda_{p}}(0)=0, \; n>0, \; \sigma=1,\ldots,p-1.
\end{equation}
This condition leads us to the following relation for the Liouville
momenta $\lambda_1,\lambda_2,\ldots,\lambda_p$:
\begin{equation}\label{NS_prim}
\lambda_{\sigma}b_{\sigma}+\lambda_{\sigma+1}b_{\sigma+1}^{-1}=0,
\end{equation}
if the modes of the $\sigma$-th current $J^{(\sigma)}(z)$ are
half-integer and
\begin{equation}\label{R_prim}
\lambda_{\sigma}b_{\sigma}+\lambda_{\sigma+1}b_{\sigma+1}^{-1}=\pm\frac{1}{2},
\end{equation}
if these modes are integer. In analogy with the representations of
the NSR  algebra let us call the primary state to be Neveu-Schwarz
with respect to the $\sigma$-th current, if (\ref{NS_prim}) holds,
and call it to be Ramond with respect to the $\sigma$-th current, if
(\ref{R_prim}) with plus or minus sign holds. In what follows we
will be interested in the representations, which are NS with respect
to all currents $J^{(\sigma)}(z)$, and in the representations, which
are R with respect to one of the currents and NS with respect to all
remaining $p-2$ currents $J^{(\sigma)}(z)$.

First we are going to consider the representation, which is NS with
respect  to all $p-1$ currents $J^{(\sigma)}(z)$. Let us use the
following notation for the primary state of this representation
\begin{equation}
V^{(1)}_{\lambda_{1}^{0}} V^{(2)}_{\lambda_{2}^{0}} \ldots
V^{(p)}_{\lambda_{p}^{0}}, \label{pr1}
\end{equation}
where the notation $\lambda_{\sigma}^{0}$ means, that these
Liouville momenta  are subject to the conditions
\begin{equation}\label{NS_representation_lambda}
    \lambda_{\sigma}^{0} b_{\sigma}+\lambda_{\sigma+1}^{0} b_{\sigma+1}^{-1}=0, \quad \sigma=1,...,p-1.
\end{equation}
Thus, we have $p-1$ equations for $p$ variables. It means, that the only
one  of these variables is independent and the representation can be
labelled by the only one variable $\lambda$ and it is convenient to
parametrize $\lambda_{\sigma}^{0}$ in the following way
\begin{equation}\label{NS_rep}
    \lambda_{\sigma}^{0}=\frac{\lambda}{\sqrt{(\sigma Q-pb)(pb-(\sigma-1)Q)}},\quad \sigma=1,...,p,
\end{equation}
which is automatically consistent with
(\ref{NS_representation_lambda}). It is easy  now to calculate the
conformal dimension of the primary state (\ref{pr1}):
\begin{equation}
\Delta_{p,0}(\lambda)=\frac{1}{p}\left(\frac{Q^2}{4}-\lambda^2
\right).
\end{equation}
The OPEs of all the currents $J^{(\sigma)}(z)$ with the corresponding primary state are given by
\begin{multline}\label{J_current_OPE_NS}
    J^{(\sigma)}(z)V^{(1)}_{\lambda_{1}^{0}}(0)... V^{(p)}_{\lambda_{p}^{0}}(0)=\sum_{m=\pm 1}
    C^{(\sigma)}(m,m;\lambda_1,\ldots,\lambda_p) \left[ V^{(1)}_{\lambda_{1}^{0}} ... V^{(\sigma)}_{\lambda_{
    \sigma}^{0}+\frac{m b_{\sigma}}{2}}
    V^{(\sigma+1)}_{\lambda_{\sigma+1}^{0}+\frac{m}{2b_{\sigma+1}}}...V^{(p)}_{\lambda_{p}^{0}} \right].
\end{multline}

Let us now consider the representations, which are R with respect to the
$s$-th  current $J^{(s)}(z)$ ($s=1,\ldots,p-1$) and NS with respect
to all other currents. We denote the $s$-th representation of this
type as
\begin{equation}
V^{(1)}_{\lambda_{1}^{s}} V^{(2)}_{\lambda_{2}^{s}} \ldots
V^{(p)}_{\lambda_{p}^{s}}, \label{pr2}
\end{equation}
where
\begin{equation}\label{R_rep}
    \lambda_{\sigma}^{s} b_{\sigma}+\lambda_{\sigma+1}^{s} b_{\sigma+1}^{-1}=\left\{
    \begin{array}{c}
      0, \;\; \sigma \neq s, \\
      -\frac{1}{2}, \sigma=s.
    \end{array}
    \right.
\end{equation}
Of course, we can take $+\frac{1}{2}$ instead of $-\frac{1}{2}$, but
this  will give us an equivalent representation. Notice, that there
exist therefore $p-1$ representations, which are R with respect to
the only one of the currents.

For the $s$-th representation we again have $p-1$ equations
(\ref{R_rep})  on $p$ variables $\lambda_{\sigma}^{s}$ and thus the
representations of this type can be parametrized by the only one
variable. There exists a convenient parametrization of
$\lambda_{\sigma}^{s}$ in terms of $\lambda_{\sigma}^{0}$
(\ref{NS_rep}), which automatically satisfies (\ref{R_rep}).
Introducing the new variables $d^s_{\sigma}$
\begin{align}
d_{\sigma}^{s}=\begin{cases}
\frac{1}{p}\sigma(p-s), \quad \textrm{if}\quad \sigma\leqslant s \\
\frac{1}{p}s(p-\sigma),\quad \textrm{if}\quad \sigma> s
\end{cases},\quad s=0,..,p-1,\quad \sigma =1,...,p  \label{ds}
\end{align}
we have for $\lambda_{\sigma}^{s}$:
\begin{equation}
\lambda_{\sigma}^{s}=\lambda_{\sigma}^{0}+d^s_{\sigma-1}
\frac{b_{\sigma}^{-1}}{2}+d^{s}_{\sigma}\frac{b_{\sigma}}{2},
\end{equation}
where $\lambda_{\sigma}^{0}$ are given by (\ref{NS_rep}). The
conformal dimension of the primary state (\ref{pr2}) is
\begin{equation}
\Delta_{p,s}(\lambda)=\frac{1}{p}\left(\frac{Q^2}{4}-\lambda^2
\right)+\frac{s(p-s)}{4p}.
\end{equation}
The OPEs of the currents $J^{(\sigma)}(z)$ with the corresponding primary state are given by
\begin{align}\label{J_current_OPE_R}
& J^{(\sigma)}(z)V^{(1)}_{\lambda_{1}^{s}}(0)...
V^{(p)}_{\lambda_{p}^{s}}(0)=\sum_{m=\pm 1}
    C^{(\sigma)}(m,m;\lambda_1,...,\lambda_p) \left[ V^{(1)}_{\lambda_{1}^{s}} ... V^{(\sigma)}_{\lambda_{
    \sigma}^{s}+\frac{m b_{\sigma}}{2}}
    V^{(\sigma+1)}_{\lambda_{\sigma+1}^{s}+\frac{m}{2b_{\sigma+1}}}...V^{(p)}_{\lambda_{p}^{s}} \right], \, \sigma \neq s, \notag \\
& J^{(s)}(z)V^{(1)}_{\lambda_{1}^{s}}(0)...
V^{(p)}_{\lambda_{p}^{s}}(0)=\sum_{m=\pm 1}z^{-\frac{m}{2}}
    C^{(\sigma)}(m,m;\lambda_1,...,\lambda_p)
    \left[ V^{(1)}_{\lambda_{1}^{s}} ... V^{(\sigma)}_{\lambda_{
    \sigma}^{s}+\frac{m b_{\sigma}}{2}}
    V^{(\sigma+1)}_{\lambda_{\sigma+1}^{s}+\frac{m}{2b_{\sigma+1}}}...V^{(p)}_{\lambda_{p}^{s}} \right].
\end{align}

Now we are going to describe the structure of the considered
representations and calculate their characters. The states in all
representations are generated by the $p$ stress tensors and $p-1$
holomorphic  currents. The structure of the OPEs of these currents
with the primary states (\ref{J_current_OPE_NS}) and
(\ref{J_current_OPE_R}) tells us that the representation besides the
Virasoro descendants of the primary state contains also the states,
whose Liouville momenta of the $\sigma$-th and $\sigma+1$-th fields
are shifted by $\pm b_{\sigma}/2$ and $\pm b_{\sigma+1}^{-1}/2$
respectively and which are also primary with respect to the $p$
stress tensors. Thus, taking the OPE of the current
$J^{(\sigma)}(z)$ with these states with shifted momenta
$\lambda_{\sigma}$ and $\lambda_{\sigma+1}$ we will generate the
infinite number of states, which are  primary with respect to the
$p$ stress tensors. For the $s$-th representation these states are
given by
\begin{equation}\label{fstate}
    V^{(1)}_{\lambda_{1}^{s}+n_1\frac{b_{1}}{2}}
    V^{(2)}_{\lambda_{2}^{s}+n_1\frac{b_{2}^{-1}}{2}+n_2\frac{b_{2}}{2}} ...
    V^{(p)}_{\lambda_{p}^{s}+n_{p-1}\frac{b_{p}^{-1}}{2}},
\end{equation}
where $n_{\sigma} \in \mathbb{Z}$ and $n_0=n_p=0$. In addition, $p$
stress tensors generate  the Virasoro submodules from each of
(\ref{fstate}). Therefore, the $s$-th representation of the first
realization of $\mathcal{A}(2,p)$, which we denote by
$\pi^{\textbf{1}}_{p,s}$ is given by the following expression
\begin{equation}\label{virrep1}
   \pi^{\textbf{1}}_{p,s}\mathop{=}^{\textrm{def}}\;  \bigoplus_{n_1,..., n_{p-1} \in \mathbb{Z}} \big[\textrm{V}^{(1)}_{\lambda_{1}^{s}+n_1\frac{b_{1}}{2}} \big] \times
    \big[\textrm{V}^{(2)}_{\lambda_{2}^{s}+n_1\frac{b_{2}^{-1}}{2}+n_2\frac{b_{2}}{2}} \big] \times ...
    \times  \big[\textrm{V}^{(p)}_{\lambda_{p}^{s}+n_{p-1}\frac{b_{p}^{-1}}{2}} \big],
\end{equation}
where $s=0,...,p-1$ and square brackets denote the Virasoro module.

The character of the  representation $\pi^{\textbf{1}}_{p,s}$ can be
now easily calculated. We have
\begin{equation}
\chi_{p}^{s}(q)=\textrm{tr} \left\{ q^{\sum_{\sigma=1}^p
L^{(\sigma)}_{0}} \right\}_{\pi^{\textbf{1}}_{p,s}},
\end{equation}
where $L^{(\sigma)}_0$ is the element of the Virasoro subalgebra
generated by the  stress tensor $T^{(\sigma)}$. Thus,
\begin{align}\label{}
    \chi_{p}^{s}(q)=(\chi_B(q))^{p} \sum_{n_1, \ldots, n_{p-1} \in \mathbb{Z} \atop n_0=n_p=0}
    q^{\sum_{\sigma=1}^{p} \Delta^{(\sigma)}(\lambda_{\sigma}^{s}+n_{\sigma-1}\frac{b_{\sigma}^{-1}}{2}+n_{\sigma}\frac{b_{\sigma}}{2})}.
\end{align}
Calculating  the sum of conformal dimensions one gets
\begin{equation}\label{}
    \chi_{p}^{s}(q)=q^{\Delta_{p,s}(\lambda)} (\chi_B(q))^{p} \sum_{n_1, \ldots, n_{p-1} \in \mathbb{Z} \atop
    n_0=n_p=0} q^{\frac{1}{2}\sum_{\sigma=1}^{p-1}
    (n_{\sigma}^2-n_{\sigma}n_{\sigma+1})+\frac{1}{2}n_s}, \label{Vircar}
\end{equation}
where $ \Delta_{p,s}(\lambda)=(Q^{2}/4-\lambda^2 )/p+s(p-s)/(4p)$,
and $\chi_B(q)$ is defined in (\ref{bosc}).

\subsection{Comparison with the generating functions of the colored Young diagrams}

As it was argued in the Introduction, the reason for the AGT relation
is the statement that one can construct the action of the symmetry
algebra $\mathcal{A}(2,p)$ on the equivariant cohomologies of the
moduli space  of instantons $\bigsqcup_{N}
\mathcal{M}(2,N)^{\mathbb{Z}_p}$. Now we obviously see from
(\ref{charcn}) and (\ref{Vircar}) the coincidence of the generating
functions of the fixed points on the moduli space and the characters
of the first realization of $\mathcal{A}(2,p)$:
\begin{align}
q^{-\Delta_{p,s}(\lambda)}(\chi_B(q))^{p}
\chi_{p}^{s}(q)=\sum_{k_{1},...,k_{p-1}=0}^{1}q^{-\frac{1}{2}\sum_{i=1}^{p-1}(k_{i}^{2}-k_{i}k_{i+1}+\frac{2k_{i}}{p})+\frac{k_{s}}{2}}
\chi_{0,s}(k_{1},...,k_{p-1}|q) , \label{character_identity}
\end{align}
where $s=0,..,p-1$ and the generating function
 $\chi_{0,s}(k_{1},...,k_{p-1})$ is given
in (\ref{charc}). Note that the obtained identity establishes the
correspondence between the characters of the representations of the
algebra $\mathcal{A}(2,p)$ in the first realization and the
generating functions of the fixed points of the moduli space
labelled by the colored Young diagrams. Below we illustrate the
obtained identity by listing the examples for $p=2,3,4$. Using the
formula (\ref{supident}) we find

The $p=2$ case:
\begin{align}
q^{-\Delta_{2,0}(\lambda)}(\chi_{B}(q))^{2}\chi_{2}^{0}(q)&=\chi_{0,0}(0|q)+q^{-1}\chi_{0,0}(1|q)  \\
q^{-\Delta_{2,1}(\lambda)}(\chi_{B}(q))^{2}\chi_{2}^{1}(q)&=
2\chi_{0,1}(0|q). \notag
\end{align}
The $p=3$ case:
\begin{align}
&q^{-\Delta_{3,0}(\lambda)}(\chi_{B}(q))^{3}\chi_{3}^{0}(q) =\chi_{0,0}(0,0|q)+3q^{-\frac{5}{6}}\chi_{0,0}(1,0|q), \\
&q^{-\Delta_{3,1}(\lambda)} (\chi_{B}(q))^{3}\chi_{3}^{1}(q)
=3\chi_{0,1}(0,0|q)+q^{-\frac{5}{6}} \chi_{0,1}(0,1|q). \notag
\end{align}
The $p=4$ case:
\begin{align}
&  q^{-\Delta_{4,0}(\lambda)}(\chi_{B}(q))^{4}\chi_{4}^{0}(q) =\chi_{0,0}(0,0,0|q)+6q^{-\frac{3}{4}}\chi_{0,0}(1,0,0|q)+q^{-\frac{3}{2}}\chi_{0,0}(1,0,1|q) , \\
&  q^{-\Delta_{4,1}(\lambda)}(\chi_{B}(q))^{4}\chi_{4}^{1}(q)= 4\chi_{0,1}(0,0,0|q)+4q^{-\frac{3}{4}}\chi_{0,1}(0,0,1|q) , \notag \\
 & q^{-\Delta_{4,2}(\lambda)} (\chi_{B}(q))^{4}\chi_{4}^{2}(q)=6\chi_{0,2}(0,0,0|q)+2q^{-\frac{3}{4}}\chi_{0,2}(1,0,0|q). \notag
\end{align}

\section{Second realization of the algebra $\mathcal{A}(2,p)$} \label{second_realization}
The present section is devoted to the other realization of
$\mathcal{A}(2,p)$ as the product of consecutive Minimal models and
coset. This relalization is depicted by the arrow 3 on the figure
(\ref{algebra_corr}). Let us again start with the definition of the
algebra $\mathcal{A}(r,p)$:
\begin{equation}
  \mathcal{A}(r,p)=
  \frac{\widehat{\mathfrak{gl}}(n)_r}{\widehat{\mathfrak{gl}}(n-p)_r}.
  \label{Arp2}
\end{equation}
The usage of the level-rank duality allows us to rewrite (\ref{Arp2}) in the following way
\begin{equation}\label{level-rank_duality}
    \mathcal{A}(r,p) \supset \mathcal{H} \times \widehat{\mathfrak{sl}}(p)_r \times \frac{\widehat{\mathfrak{sl}}(r)_p \times
    \widehat{\mathfrak{sl}}(r)_{n-p}}{\widehat{\mathfrak{sl}}(r)_n}.
\end{equation}
Utilizing the method of \cite{Crnkovic:1989ug,Lashkevich:1992sb}, i.e. formally multiplying and dividing (\ref{level-rank_duality}) by the algebras $\widehat{\mathfrak{gl}}(k)_r$ with $k=2,\ldots,p-1$ we have
\begin{equation}\label{level-rank_duality1}
    \mathcal{A}(r,p) \supset \widehat{\mathfrak{gl}}(1)_r \times \frac{\widehat{\mathfrak{gl}}(2)_r}{\widehat{\mathfrak{gl}}(1)_r}
     \times ... \times \frac{\widehat{\mathfrak{gl}}(p)_r}{\widehat{\mathfrak{gl}}(p-1)_r}
      \times \frac{\widehat{\mathfrak{sl}}(r)_p \times
    \widehat{\mathfrak{sl}}(r)_{n-p}}{\widehat{\mathfrak{sl}}(r)_n}.
\end{equation}
Then applying the level-rank duality as we did in (\ref{A_definition1}) we get
\begin{equation}\label{level-rank_duality3}
    \mathcal{A}(r,p) \supset \mathcal{H}^p \times \frac{\widehat{\mathfrak{sl}}(r)_1 \times
    \widehat{\mathfrak{sl}}(r)_{1}}{\widehat{\mathfrak{sl}}(r)_2} \times ... \times \frac{\widehat{\mathfrak{sl}}(r)_1 \times
    \widehat{\mathfrak{sl}}(r)_{p-1}}{\widehat{\mathfrak{sl}}(r)_p} \times \frac{\widehat{\mathfrak{sl}}(r)_p \times
    \widehat{\mathfrak{sl}}(r)_{n-p}}{\widehat{\mathfrak{sl}}(r)_n}.
\end{equation}
The cosets $\widehat{\mathfrak{sl}}(r)_1 \times \widehat{\mathfrak{sl}}(r)_{m-1}/\widehat{\mathfrak{sl}}(r)_m$ with $m=2,..., p$ are the consecutive Minimal models with the $\textrm{W}_r$-symmetry \cite{Lukyanov:1989gg} and  with the central charges
\begin{equation}\label{}
    c^{\textrm{MM}}_{r,m}=(r-1)\bigg( 1-\frac{r(r+1)}{(m+r-1)(m+r)} \bigg), \quad m=2,...,p,
\end{equation}
while the coset $\widehat{\mathfrak{sl}}(r)_p \times \widehat{\mathfrak{sl}}(r)_{n-p}/\widehat{\mathfrak{sl}}(r)_p$ determines the conformal field theory with the central charge
\begin{equation}\label{}
    c_{\textrm{WPF}}=\frac{p(r^2-1)}{p+r} \bigg(1-\frac{r(p+r)}{(n-p+r)(n+r)} \bigg).
\end{equation}
In the case of $\mathcal{A}(2,p)$ or $r=2$,  the cosets
$\widehat{\mathfrak{sl}}(2)_1 \times
\widehat{\mathfrak{sl}}(2)_{m-1}/\widehat{\mathfrak{sl}}(2)_m$ are
isomorphic to the symmetry algebras of the Minimal models
$\mathcal{M}(m+1/m+2)$ \cite{Goddard:1984vk,Goddard:1986ee}. And we
have\footnote{The symmetry of the $\mathbb{Z}_p$ parafermionic
Liouville field theory is described by the coset
$\widehat{\mathfrak{sl}}(2)_p
\times\widehat{\mathfrak{sl}}(2)_{n-p}/\widehat{\mathfrak{sl}}(2)_n$
\cite{Baseilhac:1998zt,Bershtein:2010wz,Wyllard:2011mn}.}
\begin{align}
\mathcal{A}(2,p) \supset \frac{\widehat{\mathfrak{sl}}(2)_p \times
\widehat{\mathfrak{sl}}(2)_{n-p}}{\widehat{\mathfrak{sl}}(2)_n}
\times\mathcal{M}(3/4) \times...\times\mathcal{M}(p+1/p+2)\times
\mathcal{H}^{p} . \label{A2psr}
\end{align}
Further we will show that the  character of a certain sum of the
representations of the right hand side of (\ref{A2psr}) coincides
with the character of the representation of the first realization
of $\mathcal{A}(2,p)$, which means that  two realizations of
$\mathcal{A}(2,p)$ are consistent. Then, automatically all
characters will be equal to the sum of the generating functions of
the pairs of colored Young diagrams.

\subsection{Representations of the
coset $\widehat{\mathfrak{sl}}(2)_p \times \widehat{\mathfrak{sl}}(2)_{n-p}/\widehat{\mathfrak{sl}}(2)_{n}$}

In this section we are going to  review the representations of the
coset $\widehat{\mathfrak{sl}}(2)_p \times
\widehat{\mathfrak{sl}}(2)_{n-p}/\widehat{\mathfrak{sl}}(2)_{n}$
\cite{Goddard:1984vk,Goddard:1986ee}.  Let us denote the integrable
representation of $\widehat{\mathfrak{sl}}(2)_p$ by
$\pi_{p,\frac{m}{2}}$, where $0 \leq m \leq p$ and the
representation of $\widehat{\mathfrak{sl}}(2)_{n-p}$ by
$\pi_{n-p,j}$, where $j$ is a continuous parameter. The
representation of the numerator $\pi_{p,\frac{m}{2}} \times
\pi_{n-p,j}$ is decomposed into the sum of the irreducible
representations of the product of the denominator and coset itself:
\begin{align}
\label{representation_decomposition} \pi_{p,\frac{m}{2}} \otimes
\pi_{n-p,j} =  \underset{s \in \mathbb{Z} \atop m-s=0 \,
\textrm{mod} \, 2}{\oplus} \pi_{n,j+\frac{s}{2}} \otimes V^m_s(p,j),
\end{align}
where $\pi_{n,j+\frac{s}{2}}$ is the representation  of the
denominator $\widehat{\mathfrak{sl}}(2)_{n}$ and $V^m_s(p,j)$ is the
representation of the coset $\widehat{\mathfrak{sl}}(2)_p \times
\widehat{\mathfrak{sl}}(2)_{n-p}/\widehat{\mathfrak{sl}}(2)_{n}$.

The characters  $c_{s}^{m}(q)$ of the representations of
$\widehat{\mathfrak{sl}}(2)_p \times
\widehat{\mathfrak{sl}}(2)_{n-p}/\widehat{\mathfrak{sl}}(2)_{n}$ are
given by the branching functions which can be found from the
relation for the characters originating from
(\ref{representation_decomposition}):
\begin{equation}
\chi^{\mathfrak{sl}(2)}_{p,\frac{m}{2}}(q,z)\chi^{\mathfrak{sl}(2)}_{n-p,j}(q,z)
=\sum_{s \in \mathbb{Z} \atop m-s=0 \, \textrm{mod} \, 2}
\chi^{\mathfrak{sl}(2)}_{n,j+\frac{s}{2}}(q,z) c^m_s(q).
\end{equation}
These characters are labelled by the integer parameters $m$ and $s$
and  continuous parameter $j$ and are given by
\cite{Kakushadze:1993jf}:
\begin{multline}
c^{m}_{s}(q)=q^{\delta^{m}_{s}(j)} \chi_B^3(q) \sum_{r,l=0}^{+\infty} (-1)^{r+l}
q^{\frac{l(l+1)}{2}+\frac{r(r+1)}{2}+rl(p+1)} \times \\
\times \left(q^{l\frac{m-s}{2}+r\frac{m+s}{2}}-q^{p+1-m+l(p+1-\frac{m-s}{2})
+r(p+1-\frac{m+s}{2})} \right), \label{coset_character}
\end{multline}
where $0 \leqslant m \leqslant p, \; m-s=0 \, \textrm{mod} \, 2,$ and
\begin{equation}\label{pfcoset_dimension}
\delta^{m}_{s}(j)=\frac{j(j+1)}{n-p+2}+\frac{m(m+2)}{4(p+2)}-\frac{(2j+s)(2j+s+2)}{4(n+2)}.
\end{equation}
Using the formula (\ref{coset_character}) it can be shown that $c^{m}_{s}(q)$ admits the following symmetries
\begin{align}\label{character_symmetries}
& c^m_{-s}(q)=c^m_s(q), \\
& c^m_{s+2p}(q)=c^m_s(q). \notag
\end{align}
Thus, we can confine our consideration to $s$, which are integers
from $0$ to $p$.

Below we are going to pass to the different parametrization of  the
highest weight state and of the character of the highest weight
representation, which is convenient in our calculations. So we pass
from the  parameters $n$ and $j$ to the $b$ and $\mu$ as follows
\begin{align}\label{j_parametrization}
b^2=-\frac{n+2}{n-p+2} ,\quad j=\frac{1}{Q}\left(\mu-\frac{Q}{2}-\frac{s}{2b} \right), \quad Q=b+b^{-1}.
\end{align}
And we denote the highest weight  state of the coset by
$\Psi_{s}^{m}(\mu)$ and  its representation by
$[\Psi_{s}^{m}(\mu)]$. The dimension of the highest weight state and
the character of this representation are given by
\begin{align}
& \Delta^m_s(\mu)=
\begin{cases}
\frac{1}{p}\left( \frac{Q^2}{4}-\mu^2 \right)+\frac{s(p-s)}{2p(p+2)}+\frac{(m-s)(m+s+2)}{4(p+2)}, \; m \geq s \\
\frac{1}{p}\left( \frac{Q^2}{4}-\mu^2 \right)+\frac{s(p-s)}{2p(p+2)}+\frac{(s-m)(2p-m-s+2)}{4(p+2)}, \; m<s
\end{cases}  \label{dimcos} \\
& c_{s}^{m}(q)=q^{D_{s}^{m}(\mu)} \chi^{3}_{B}(q) \sum_{r,l=0}^{\infty}(-1)^{r+l}q^{\frac{l(l+1)}{2}+\frac{r(r+1)}{2}+r l(p+1)}\times \notag \\
& \qquad\qquad\qquad \times (q^{l\frac{m-s}{2}+r\frac{m+s}{2}}-q^{p+1-m+l(p+1-\frac{m-s}{2})+r(p+1-\frac{m+s}{2})}). \label{charchar}
\end{align}
where $0 \leqslant m,s \leqslant p, \; m-s=0 \, \textrm{mod} \, 2$ and

\begin{equation*}
D^m_s(\mu)=\frac{1}{p}\left( \frac{Q^2}{4}-\mu^2 \right)+\frac{s(p-s)}{2p(p+2)}+\frac{(m-s)(m+s+2)}{4(p+2)}
\end{equation*}
for all $m$ and $s$.

\subsection{Product of consecutive Minimal models}

The other main part of (\ref{A2psr})  is the product of the Minimal
models. Minimal model $\mathcal{M}(m/m+1)$ has the central charge
$c_{2,m}^{\textrm{MM}} = 1- 6/(m(m+1))$   and has the finite set of
primary fields $\phi^{(m)}_{r,s}$ with $r=1,2,...m-1$, and
$s=1,2,...m$  \cite{Belavin:1984vu}. The  dimensions of these
primary fields are given by the formula
\begin{align}
h_{r,s}^{(m)} = \frac{((m+1)r-ms)^{2}-1}{4m(m+1)}.
\end{align}
The following fields are identified with each other
$\phi^{(m)}_{r,s} = \phi^{(m)}_{m-r,m+1-s}$. We will  denote the
irreducible Virasoro representation, built from the highest weight
state $\phi_{r,s}^{(m)}$ by $M^{(m)}_{r,s}$. The character of such a
representation equals to
\begin{align}
\chi_{r,s}^{(m)}(q)=\left.\textrm{Tr}(q^{L_{0}})\right|_{M^{(m)}_{r,s}}=  \Delta_{r,s}^{m}(q) \chi_{B}(q),
\end{align}
where
\begin{align}
&\Delta_{r,s}^{m}(q)  = \sum_{k\in
\mathbb{Z}}(q^{\alpha_{r,s}^{m}(k)}-q^{\alpha_{r,-s}^{m}(k)}), \quad
\alpha_{r,s}^{m}(k)= \frac{(2m(m+1)k-s m+r(m+1))^{2}-1}{4m(m+1)}.
\end{align}

Consider the product of $p-1$ Minimal models highest weight states $
\phi_{1k_{1}}^{(3)} \times \phi_{k_{1}k_{2}}^{(4)} \times...\times
\phi_{k_{p-2}n}^{(p+1)}$, where $k_{i}$ runs from $1$ to $i+2$ and
$n$ runs from $1$ to $p+1$. This composite highest weight state has
the dimension (hereafter we imply $k_{0}=1, k_{p-1}=n$)
\begin{align}
& h_{n}(k_{1},...,k_{p-2})=\sum_{i=1}^{p-1}h^{(i+2)}_{k_{i-1}k_{i}}= \frac{(n^{2}-1)(p+1)}{4(p+2)}
+\frac{1}{2} \sum_{i=0}^{p-2}(k_{i}^{2}-k_{i}k_{i+1}). \label{hn}
\end{align}
The irreducible representation which is built from this composite
highest weight state  is denoted as $M^{(3)}_{1,k_{1}} \times
M^{(4)}_{k_{1},k_{2}}\times ....\times M^{(p+1)}_{k_{p-2},n}$. Below
we will consider the following sum of the representations
\begin{equation}
\bigoplus_{\substack{\{k_{1},..,k_{p-2}\}\\
1 \leq k_{i} \leq i+2}} M^{(3)}_{1,k_{1}} \times M^{(4)}_{k_{1},k_{2}}\times ....\times M^{(p+1)}_{k_{p-2},n}
\end{equation}
Denote a character of this sum of the representations by
\begin{align}
\textrm{ch}_{n}(q) \mathop{=}^{\textrm{def}}\sum_{\substack{\{k_{1},..,k_{p-2}\}
\\ 1\leq k_{i}\leq i+2, k_{p-1}=n}}\prod_{i=1}^{p-1}\chi^{(i+2)}_{k_{i-1}k_{i}}(q). \label{chch}
\end{align}
In the next subsection we will show that the characters of the
product of consecutive Minimal models and the coset coincide with
the characters of the first realization of $\mathcal{A}(2,p)$.

\subsection{Comparison with the first realization of $\mathcal{A}(2,p)$}

Let us start by considering the following sum of the representations
of the coset and consecutive Minimal models
\begin{equation}\label{rep_2nd_real}
  [\Psi^m_{s}(\mu)] \times  \bigoplus_{\substack{\{k_{1},..,k_{p-2}\}\\
1\leqslant k_{i}\leqslant
i+2}} M^{(3)}_{1,k_{1}} \times M^{(4)}_{k_{1},k_{2}}\times ....\times M^{(p+1)}_{k_{p-2},n},
\end{equation}
where $1 \leqslant n \leqslant p+1$, $0 \leqslant m,s \leqslant p$
with $m-s =0 \; \textrm{mod} \; 2$.  The character of such a
representation is equal to
\begin{equation}\label{coset_min_mod_character}
c^m_s(q)\textrm{ch}_n(q).
\end{equation}
Note, that the representation (\ref{rep_2nd_real}) is labelled by
three integer  parameters $0 \leq m,s \leq p$, $m-s=0 \;
\textrm{mod} \; 2$, $1 \leq n \leq p+1$ and one continuous parameter
$\mu$. Remember that we are looking for the sum of the
representations of the form (\ref{rep_2nd_real}), character of which
coincides with the character $\chi_{p}^{s}(q)$ of the representation
of the first realization of $\mathcal{A}(2,p)$, which is labelled by
the integer parameter $0 \leq s \leq p-1$ and continuous parameter
$\lambda$.

The character (\ref{Vircar}) of the representation of $p$ models
$\chi_{p}^{s}(q)$  looks like $q^{\Delta_{p,s}(\lambda)}$ multiplied
by some series containing integer and half-integer powers of $q$ and
the character (\ref{coset_min_mod_character}) is the sum over
$k_i$'s of $q^{\Delta^m_s(\lambda)+h_n(k_1,\ldots,k_{p-2})}$
multiplied by the series containing integer powers. Because
$h_n(k_1,\ldots,k_{p-2})-h_n(k'_1,\dots,k'_{p-2}) \in \mathbb{Z}/2$,
the necessary condition for the characters to coincide will be
\begin{equation}
\Delta^m_s(\mu)+h_n(k_1,...,k_{p-2})-\Delta_{p,s}(\lambda) \in
\mathbb{Z}/2.
\end{equation}
Performing some algebra we get
\begin{equation}
\frac{\lambda^2-\mu^2}{4p}+\frac{(m+1)^2-n^2}{4(p+2)}-\frac{s}{4}
+\frac{1}{2} \sum_{i=0}^{p-2}(k_{i}^{2}-k_{i}k_{i+1}) \in
\mathbb{Z}/2.
\end{equation}
Because the left hand side of the  expression must be half integer
and must not depend on any continuous parameter it is natural to
assume $\mu=\lambda$. This leads to the following Diophantine
equation
\begin{equation}
\frac{(m+1)^2-n^2}{4(p+2)}-\frac{s}{4}+\frac{1}{2} \sum_{i=0}^{p-2}(k_{i}^{2}-k_{i}k_{i+1}) \in \mathbb{Z}/2,
\end{equation}
from which it follows that
\begin{equation}
\frac{(m-n+1)(m+n+1)}{p+2} \in \mathbb{Z}.
\end{equation}
Because $-p \leqslant m-n+1 \leqslant p$ and $2 \leqslant m+n+1 \leqslant p+2$, there exist two possibilities
\begin{align}
& n=m+1, \quad n=p-m+1. \label{solution_nec_cond}
\end{align}
Therefore, taking into account (\ref{solution_nec_cond}), we
conclude,  that we should take only the representations of the coset
and product of Minimal models,  which have the form\footnote{These
two solutions (\ref{solution_nec_cond}) for $n$ lead to the
following equations for $k_i$'s
\begin{align*}
\sum_{i=0}^{p-2}(k_{i+1}-k_{i})^2 =
\begin{cases}
s, \;\;\;\qquad\qquad \textrm{if}\quad n=m+1\\
p+s-2m \quad \textrm{if}\quad n=p-m+1.
\end{cases}
\end{align*}}
\begin{align}
 & [\Psi^m_{s}(\lambda)]  \times \bigoplus_{\substack{\{k_{1},..,k_{p-2}\}\\
1\leqslant k_{i}\leqslant
i+2}} M^{(3)}_{1,k_{1}} \times M^{(4)}_{k_{1},k_{2}}\times ....\times M^{(p+1)}_{k_{p-2},m+1},   \notag \\
& [\Psi^m_{s}(\lambda)] \times \bigoplus_{\substack{\{k_{1},..,k_{p-2}\}\\
1\leqslant k_{i}\leqslant i+2}} M^{(3)}_{1,k_{1}} \times
M^{(4)}_{k_{1},k_{2}}\times ....\times M^{(p+1)}_{k_{p-2},p-m+1} .
\label{repscos}
\end{align}
Therefore the representation of the algebra $\mathcal{A}(2,p)$ in
the second realization is
\begin{align}
\pi_{p,s}^{\textbf{2}}=\bigoplus_{0 \leqslant m \leqslant p \atop
m-s=0
\; \textrm{mod} \; 2}[\Psi^m_{s}(\lambda)]  \times \bigoplus_{\substack{\{k_{1},..,k_{p-2}\}\\
1\leqslant k_{i}\leqslant i+2}} M^{(3)}_{1,k_{1}} \times
M^{(4)}_{k_{1},k_{2}}\times ....\times (M^{(p+1)}_{k_{p-2},m+1}
\oplus M^{(p+1)}_{k_{p-2},p-m+1}).
\end{align}
The character of this representation is
\begin{equation}\label{main_hypothesis0}
\sum_{0 \leqslant m \leqslant p \atop m-s=0 \; \textrm{mod} \; 2}
c^m_s(q) (\textrm{ch}_{m+1}(q) +\textrm{ch}_{p-m+1}(q)).
\end{equation}
The conjecture that the representation $\pi_{p,s}^{\textbf{2}}$ is
the another form of representation of $\mathcal{A}(2,p)$, i.e
\begin{align}
\pi_{p,s}^{\textbf{1}}\cong \pi_{p,s}^{\textbf{2}},
\end{align}
leads us to the following non-trivial identity:
\begin{equation}\label{main_hypothesis}
\sum_{0 \leqslant m \leqslant p \atop m-s=0 \; \textrm{mod} \; 2}
c^m_s(q) (\textrm{ch}_{m+1}(q)
+\textrm{ch}_{p-m+1}(q))=q^{\Delta_{p,s}(\lambda)} (\chi_B(q))^{p}
\sum_{n_1, \ldots, n_{p-1} \in \mathbb{Z} \atop
    n_0=n_p=0} q^{\frac{1}{2}\sum_{\sigma=1}^{p-1}
    (n_{\sigma}^2-n_{\sigma}n_{\sigma+1})+\frac{1}{2}n_s},
\end{equation}
where expressions for the characters are given in   (\ref{Vircar}),
(\ref{charchar}), (\ref{chch}). We have checked the equality
(\ref{main_hypothesis}) for the cases $p=2,...,8$  order by order up
to $q^6$. The equality (\ref{main_hypothesis}) proves the
consistence of the representations of the algebra $\mathcal{A}(2,p)$
in the first and second realizations. It should be noted that in \cite{Belavin:2012bd} it was shown, that for the case when the equivariant parameter $n=1$ (and in principal for all $n \in \mathbb{Z}$), the characters of the representations of the second realization of $\mathcal{A}(2,p)$ can be rewritten as certain series using Generalized Rogers-Ramanujan identities.

\section{Comparison of the instanton partition functions} \label{partition_functions}

There exist two approaches to the calculation of the instanton
partition functions in the $\mathcal{N}=2$ supersymmetric gauge
theory on $\mathbb{C}^2/\mathbb{Z}_p$. The difference in these
approaches is in the compactification of the instanton moduli space.
As the result one gets different expressions for the same
instanton partition function. The first compactification of the moduli space is related to the first realization of the algebra $\mathcal{A}(2,p)$. However, unfortunately, we are not aware of any geometrical interpretation for the second realization of $\mathcal{A}(2,p)$.

\subsection{First compactification}

One approach to the calculation of the instanton partition function
on $\mathbb{C}^2/\mathbb{Z}_p$ is the integration over the moduli
space of instantons $\bigsqcup_N \mathcal{M}(X_p,r,N)$ on the
resolved space $X_p=\widetilde{\mathbb{C}^2/\mathbb{Z}_p}$. The
partition function in this approach was calculated in
\cite{Bonelli:2011kv,Bonelli:2012ny}:
\begin{align}
Z_{\textrm{inst}}^{p,s}(\vec{a},\epsilon_{1},\epsilon_{2}|\Lambda) = \sum_{n_1, \ldots, n_{p-1} \in \mathbb{Z} \atop
    n_0=n_p=0}
\frac{\Lambda^{(n_{i}+d_{i}^{s})C_{ij}(n_{j}+d_{j}^{s})}}{l_{p,s}^{\textrm{vec}}(a,n_{1},...,n_{p-1})}
\prod_{\sigma=1}^{p}Z_{\textrm{inst}}^{\mathbb{C}^{2}}(\vec{a}_{s}^{(\sigma)},\epsilon_{1}^{(\sigma)},\epsilon_{2}^{(\sigma)}|\Lambda), \quad s=0,...,p-1,
\label{Zonerep}
\end{align}
where $\vec{a}=(a,-a)$, $\vec{a}^{(\sigma)}_{s}=
(a^{(\sigma)}_{s},-a^{(\sigma)}_{s})$, and   $a^{(\sigma)}_{s} = a +
(n_{\sigma+1}+d_{\sigma+1}^{s})\epsilon_{1}^{(\sigma)} +
(n_{\sigma}+d_{\sigma}^{s})\epsilon_{2}^{(\sigma)}$,  and
regularization parameters are
$\epsilon_{1}^{(\sigma)}=(p-\sigma)\epsilon_{1}-\sigma\epsilon_{2}$,
$\epsilon_{2}^{(\sigma)}=(\sigma+1-p)\epsilon_{1}+(\sigma+1)\epsilon_{2}$.
The shifts $d_{\sigma}^{s}$ are given by the formula (\ref{ds}):
\begin{align}
d_{\sigma}^{s} =
\begin{cases}
\frac{1}{p}\sigma(p-s), \quad \textrm{if}\quad \sigma\leqslant s \\
\frac{1}{p}s(p-\sigma),\quad \textrm{if}\quad \sigma> s
\end{cases},\quad s=0,..,p-1,\quad \sigma =1,...,p.
\end{align}
And $C_{ij}$ is the $(p-1)\times(p-1)$ Cartan matrix of the simple Lie algebra $A_{p-1}$.

The $SU(2)$ instanton partition function on  $\mathbb{C}^{2}$ was
calculated in \cite{Nekrasov:2002qd}:
\begin{align}
Z_{\textrm{inst}}^{\mathbb{C}^{2}}(\vec{a},\epsilon_{1},\epsilon_{2}|\Lambda)
= \sum_{(Y_{1},Y_{2})} \Lambda^{|Y_{1}|+|Y_{2}|} \prod_{i,j=1}^{2} \prod_{s\in Y_{i}} \frac{1}{E_{Y_{i},Y_{j}}(s|a_{i}-a_{j})
(\epsilon_{1}+\epsilon_{2}-E_{Y_{i},Y_{j}}(s|a_{i}-a_{j}))} ,
\label{partfunc1}
\end{align}
where $|Y|$ is the total number of the boxes in the Young diagram $Y$, $s$ denotes a box in the Young diagram $Y$, and
\begin{align}
E_{Y,W}(a|s)= a - l_{W}(s)\epsilon_{1}+(a_{Y}(s)+1)\epsilon_{2},
\end{align}
where $a_{Y}(s)$ and $l_{Y}(s)$ is the arm and leg length
respectively, i.e. the number of boxes in $Y$ to the right and below
of the box $s \in Y$.

The functions $l_{p,s}^{\textrm{vec}}(a,n_{1},...,n_{p-1})$ are called blow-up factors and were calculated by geometrical methods in \cite{Bonelli:2012ny} and are given by
\begin{align}
l_{p,s}^{\textrm{vec}}(a,n_{1},...,n_{p-1}) =& \prod_{\sigma =0}^{p-1} g^{(\sigma)}(2a_{s}^{(\sigma)}, \epsilon_{1}^{(\sigma)},
\epsilon_{2}^{(\sigma)}, n_{\sigma}+d^{s}_{\sigma}, n_{\sigma+1}+d_{\sigma+1}^{s}) \times \notag \\
& \times g^{(\sigma)}(-2a_{s}^{(\sigma)}, \epsilon_{1}^{(\sigma)},
\epsilon_{2}^{(\sigma)}, -(n_{\sigma}+d^{s}_{\sigma}), -(n_{\sigma+1}+d_{\sigma+1}^{s})),
\end{align}
where
\begin{align}
g^{(\sigma)}(a,e_{1},e_{2}, \mu,\nu) = \begin{cases}
\prod\limits_{\substack{m \geqslant 0, n \leqslant -1 \\ \sigma(\nu+m) \leqslant (\sigma+1)(\mu+n)}} (a+m e_{1} +n e_{2}) , \quad \textrm{if} \quad \sigma \nu < (\sigma+1)\mu
\\
\qquad\quad\; 1, \qquad\qquad \qquad\qquad   \qquad\quad\; \textrm{if} \quad \sigma \nu = (\sigma+1)\mu \\
\prod\limits_{\substack{m \leqslant -1, n \geqslant 0 \\
\sigma(\nu+m) > (\sigma+1)(\mu+n)}} (a+m e_{1} +n e_{2}) ,  \quad
\textrm{if} \quad  \sigma \nu > (\sigma+1)\mu.
\end{cases}
\end{align}

\subsection{Second compactification}

The other compactification of the instanton moduli space is obtained
by the lift of the action of $\mathbb{Z}_p$ group in
$\mathbb{C}^2/\mathbb{Z}_p$ to the moduli space $\bigsqcup_N
\mathcal{M}(2,N)$ on $\mathbb{C}^2$. The resulting moduli space is
denoted by $\bigsqcup_N \mathcal{M}(2,N)^{\mathbb{Z}_p}$ and its
fixed points are labelled by the pairs of Young diagrams with $p$
colors. Thus, in the instanton partition function corresponding to
this compactification of the moduli space we take the sum only over these
Young diagrams, and also count only the special boxes of these Young
diagrams. So as in the Section \ref{fixed_points_counting} we take
the sum over set $\lozenge$ of the pairs of Young diagrams
$(Y_{1},Y_{2})$:
\begin{align}
\lozenge =\{(Y_{1}, Y_{2})|\;
\begin{ytableau}
r_1 & & & \\
& & \\
\
\end{ytableau}
,
\begin{ytableau}
r_2 & & \\
& \\
\
\end{ytableau},
\sharp(
\begin{ytableau}
m
\end{ytableau})-\sharp(
\begin{ytableau}
0
\end{ytableau})=k_{m}\},  \label{loze}
\end{align}
where the  box in $Y_{1}$ with the coordinates $(i,j)$ has the color
$r_{1}+i-j \, \textrm{mod} \, p$ and the box $(i,j)$ in $Y_{2}$ has
the color $r_{2}+i-j \, \textrm{mod} \, p$ and
$ \sharp (\begin{ytableau} m
\end{ytableau}),\sharp  (
\begin{ytableau}
0
\end{ytableau})$
-- the numbers of the boxes in $(Y_{1},Y_{2})$ with $m$ and $0$ color respectively.

Also introduce the formula
\cite{Fucito:2004ry,Fucito:2006kn,Flume:2002az}
\begin{multline}
\mathcal{Z}_{r_{1},r_{2}}(k_{1},...,k_{p-1}|\vec{a},\epsilon_{1},\epsilon_{2}|\Lambda)= \\
=\sum_{(Y_{1},Y_{2})\in \lozenge} \Lambda^{\frac{|Y_{1}|+|Y_{2}|}{p}} \prod_{i,j=1}^{2}
\tilde{\prod_{s\in Y_{i}}} \frac{1}{E_{Y_{i},Y_{j}}(s|a_{i}-a_{j})(\epsilon_{1}+\epsilon_{2}-E_{Y_{i},Y_{j}}(s|a_{i}-a_{j}))},
\end{multline}
where the product $\tilde{\prod}$  goes only through $s\in Y_{i}$ that
satisfy  $l_{Y_{j}}(s)+a_{Y_{i}}(s)+1 \equiv r_{j}-r_{i} \;
\textrm{mod} \; p$. After all the notations being introduced we can
present the expression for the instanton  partition function in the
second compactification:
\begin{align}
&Z_{\textrm{inst}}^{p,s}(\vec{a},\epsilon_{1},\epsilon_{2}|\Lambda)=
\sum_{k_{1},...,k_{p-1}=0}^{1}\Lambda^{-\frac{1}{2}\sum_{i=1}^{p-1}(k_{i}^{2}-k_{i}k_{i+1}+\frac{2k_{i}}{p})+\frac{k_{s}}{2}}
\mathcal{Z}_{0,s}(k_{1},...,k_{p-1}|\vec{a},\epsilon_{1},\epsilon_{2}|\Lambda).
\label{Ztworep1}
\end{align}
The two expressions for the instanton partition functions (\ref{Zonerep}) and (\ref{Ztworep1}) coincide, as it was checked in \cite{Bonelli:2012ny,Ito:2013kpa}. In the next subsection we give
 arguments in favor of this equality from the conformal field
theory point of view.

\subsection{Bases in conformal field theories and the equality of instanton partition functions}

For the cases $r=2$, $p=1$ \cite{Gaiotto:2009ma, Marshakov:2009gn} and $r=2$, $p=2$ \cite{Belavin:2011pp} it was shown that the instanton partition function of the $\mathcal{N}=2$ supersymmetric gauge theory without matter is equal to the norm of the Whittaker vector. In the mentioned cases $r=2$ and $p=1,2$ this Whittaker vector is determined as the eigenvector of the upper nilpotent subalgebra of the symmetry algebra (Virasoro in $p=1$ case and Neveu-Schwarz-Ramond in $p=2$ case). In a situation with arbitrary $p$ and $r=2$ the analogue of the Virasoro and NSR algebras is the coset

\begin{equation}\label{Whit_gen}
\frac{\widehat{\mathfrak{sl}}(2)_p \times \widehat{\mathfrak{sl}}(2)_{n-p}}{\widehat{\mathfrak{sl}}(2)_n}.
\end{equation}
Thus, let us assume that for arbitrary $p$ the Whittaker vector is the eigenvector of the upper nilpotent part of the coset (\ref{Whit_gen}). Note, that we suppose the remaining part of the $\mathcal{A}(2,p)$ algebra to act by zero on this Whittaker vector. Thus, we assume that for general $p$ we can represent partition function as the norm of the Whittaker vector $|W\rangle$
\begin{align}
Z_{\textrm{inst}} = \langle W|W\rangle.
\end{align}

As it was mentioned in Introduction, there exist two ways to
construct the moduli space of instantons for the $\mathcal{N}=2$
supersymmetric $U(r)$ gauge theory on $\mathbb{C}^2/\mathbb{Z}_p$.
And for each way of compactification of the moduli space there is a
basis of geometrical origin, which is in one-to-one correspondence
with the fixed points of the torus action. Thus, to calculate the
instanton partition function we can use the basis labelled by the
colored Young diagram, which was constructed explicitly in
\cite{Belavin:2012eg} for the cases $r=1,2$ and $p=2$, or labelled
by the $p$ $r$-tuples of ordinary Young diagrams, which was
constructed explicitly in \cite{Alba:2010qc,Belavin:2011pp} for
$r=2$ and $p=1,2$. Despite for $r=2$ and arbitrary $p$ the basis for
both compactifications of the moduli space was not constructed, we
assume its existence.

Inserting the complete set of states in the norm of the Whittaker vector in
each basis we can establish the equality between the  instanton
partition functions for the pure gauge theory calculated for the
different compactifications of the moduli space. Note that we have
already established the correspondence between the fixed points of
the moduli space in different compactifications, or, equivalently,
between the two bases of geometrical origin corresponding to these
compactifications. The form of this correspondence is given by the
identity (\ref{character_identity}). Thus, taking the basis vectors
corresponding to the fixed points labelled by the Young diagrams
with $p$ colors from one side and taking the basis vectors
corresponding to the fixed points in another compactification
labelled by the $p$ $r$-tuples of Young diagrams and $p-1$
$r$-dimensional vectors from the other side, after inserting the
full set of states in the norm of the Whittaker vector we obtain the
formula connecting the instanton partition functions in different
compactifications.

\section*{Acknowledgements}

The authors are grateful to M.Bershtein, B.Feigin, D.Gepner,
L.Spodyneiko,  A.Zamolodchikov for useful discussions and interest
to our work. M.A. thanks the Simons Center for Geometry and Physics
for the kind hospitality during the early stage of this project.

This work was supported by RFBR grants No.12-01-00836-a, 12-02-01092-a, 12-01-31236-mol a, 12-02-33011-mol a ved and by the Russian Ministry of Education and Science under the grants 2012-1.5-12-000-1011-012, contract No.8528, 2012-1.1-12-000-1011-016, contract No.8410 and agreement \newline No.14.A18.21.2027. The work of M.A. and G.T. were also supported by the 2012 Dynasty Foundation Grant and Federal Targeted Programs of the Russian Ministry of Education and Science. The research leading to these results has received funding from the
People Programme (Marie Curie Actions) of the European Union's Seventh
Framework Programme FP7/2007-2013/ under REA Grant Agreement No 317089.

\Appendix
\section{Symmetries of the generating functions} \label{app_A}

The aim of the present Appendix is to analyze the symmetries of the
generating function  of Young diagrams
$\chi_{0,s}(k_1,...,k_{p-1}|q)$ with $s=0,1,\ldots,p-1$ and $k_i$
equal to 0 or 1. Let us remember the formula (\ref{charc}) for the
generating function and write it in a more convenient form for the
further considerations
\begin{multline} \label{gen_func_app}
\chi_{0,s}(k_1,...,k_{p-1}|q)=(\chi_B(q))^{2p}q^{\frac{1}{2}\sum_{i=1}^{p-1}(k_i^2-k_i k_{i+1}+\frac{2k_i}{p})-\frac{1}{2}k_{s}} \times \\
\times \sum_{\{m_i \} \in \mathbb{Z}} q^{\frac{1}{4}\sum_{i=0}^{s-1}(2m_{i+1}-2m_{i}-k_{i+1}+k_{i}+1)^2+\frac{1}{4}\sum_{i=s-1}^{p-1}(2m_{i+1}-2m_{i}-k_{i+1}+k_{i})^2-\frac{s}{4}}.
\end{multline}
Taking some $j \neq s$ and assuming $k_j=0$ and $k_{j-1}=0$,
$k_{j+1}=1$, we make the  substitution of the summation variable
$m_j$ in (\ref{gen_func_app})
\begin{equation} \label{subst}
m_j=m_{j+1}+m_{j-1}-\tilde{m}_{j},
\end{equation}
which effectively leads to $k_j=0 \rightarrow k_j=1$. After some calculations, we have
\begin{equation}\label{char_symmetry_1}
\chi_{0,s}(...,\stackrel{j}{0,0,1},...|q)=q^{-\frac{1}{p}}\chi_{0,s}(...,\stackrel{j}{0,1,1},...|q), \; j \neq s.
\end{equation}
The same substitution as in (\ref{subst}) proves that
\begin{equation}\label{char_symmetry_2}
\chi_{0,s}(...,\stackrel{j}{1,0,0},...|q)=q^{-\frac{1}{p}}\chi_{0,s}(...,\stackrel{j}{1,1,0},...|q), \; j \neq s.
\end{equation}
Next we have to consider the situation when $k_s=0$. Assuming that
$k_{s-1}=0$ and $k_{s+1}=0$,  we make the substitution for the
summation variable $m_s$ in (\ref{gen_func_app})
\begin{equation}
m_s=m_{s+1}+m_{s-1}-\tilde{m}_s
\end{equation}
which effectively leads to $k_s=0 \rightarrow k_s=1$. After some calculations, we have
\begin{equation}\label{char_symmetry_3}
\chi_{0,s}(...,\stackrel{s}{0,0,0},...|q)=q^{-\frac{1}{p}}\chi_{0,s}(...,\stackrel{s}{0,1,0},...|q).
\end{equation}
If $k_{s-1}=1$ and $k_{s+1}=1$ the suitable substitution would be
\begin{equation}
m_s=m_{s+1}+m_{s-1}-\tilde{m}_s-1
\end{equation}
which again effectively leads to $k_s=0 \rightarrow k_s=1$. After some calculations, we have
\begin{equation}\label{char_symmetry_4}
\chi_{0,s}(...,\stackrel{s}{1,0,1},...|q)=q^{1-\frac{1}{p}}\chi_{0,s}(...,\stackrel{s}{1,1,1},...|q).
\end{equation}

The next task is to determine the classes of inequivalent generating functions for each $s$ with $k_i$ equal to 0 or 1.

Let us start from the case $s=0$. As we remember, the array of $k_i$
is a series of $p-1$  zeros and unities. We can look at this array
as on the islands of unities in the sea of zeros. Then, one can
easily see that the symmetries (\ref{char_symmetry_1}) and (\ref{char_symmetry_2}) forbid the islands to merge (at least
one 0 must be between them), but allows them to change their size.
Thus, the class of equivalence is determined by the number of
islands $n$, which takes the values $0,1,2,\ldots,\left[\frac{p}{2}
\right]$. It is convenient to choose the following representative of
the $n$-th class
\begin{equation}
\chi_{0,0}(1,0,1,0,...,1,\stackrel{2n-1}{0,1,0},...,0|q).
\end{equation}

Thus, for $s=0$ the cardinality of the $n$-th class is equal to
$\binom{p}{2n}$ (the  number of ways to distribute $2n$ borders of
the islands between $p$ places).

Now we proceed with the same calculation for the case $s > 0$.
Looking again onto the  array of $k_i$, which consists of 0 and 1,
we notice, that if there is an island of identities, containing the
position number $s$, we can destroy this island obtaining the
generating function equivalent to $\chi_{0,s}$. Then, due to the
symmetries (\ref{char_symmetry_3}) and (\ref{char_symmetry_4}) we can annihilate the islands
to left of the position $s$ with the islands to the right of the
position $s$. This means, that at the end we will be left with some
number of the islands on the one side (left or right) only. This
lead us to the conclusion that the class of equivalence in this case
is determined by the difference of the number of islands to the left
and to the right of the position $s$. Therefore, the number of the
classes of equivalence is equal to
\begin{equation}
\left[ \frac{s}{2} \right]+\left[ \frac{p-s}{2} \right]+1.
\end{equation}
Let $l$ be the difference of the number of islands to the left and
to the right of $s$-th position.  Then, the number of the generating
functions in the corresponding class of equivalence with $k_s=0$ is
given by
\begin{equation}
\sum_{j=0}^{\left[\frac{s}{2} \right]-l} \binom{s}{2n+2j} \binom{p-s}{2j}
\end{equation}

The number of the generating functions in the same class with
$k_s=1$ (which effectively leads to  the addition of one island
border on the each side) is given by
\begin{equation}
\sum_{j=0}^{\left[\frac{s}{2} \right]-l} \binom{s}{2l+2j+1}\binom{p-s}{2j+1}   .
\end{equation}
Summing up both contributions, we obtain using the Vandermonde's identity
\begin{equation}
\sum_{j=0}^{\left[\frac{s}{2} \right]-n} (\binom{s}{2l+2j} \binom{p-s}{2j}+\binom{s}{2l+2j+1} \binom{p-s}{2j+1} )=\binom{p}{s-2l}.
\end{equation}
A convenient choice of the representative of the $l$-th class is
\begin{equation}
\chi_{0,s}(0,...,\stackrel{s-2l+1}{0,1,0},1,0,...,1,0,\stackrel{s}{1,0,0},...,0).
\end{equation}

The situation is the same when we have $n$ islands on the right
side, except for we should replace $s$ by $p-s$, which gives the
cardinality $\binom{p}{p-s-2n}$. A convenient representative would
be
\begin{equation}
\chi_{0,s}(0,...,\stackrel{s}{0,0,1},0,1,...0,1,\stackrel{s+2n-1}{0,1,0},...,0).
\end{equation}

\section{Conformal field theories based on the coset}\label{app_cosets}

In the present Appendix we give some information about the conformal field theories based on the coset
\begin{equation}\label{cft_coset}
\frac{\widehat{\mathfrak{sl}}(r)_{l_1} \times\widehat{\mathfrak{sl}}(r)_{l_2}}{\widehat{\mathfrak{sl}}(r)_{l_1+l_2}}.
\end{equation}
In the case of general integer $r \geq 2$ and arbitrary complex $l_1$ and $l_2$ the coset (\ref{cft_coset}) describes the conformal field theory with the central charge
\begin{equation}
c(r,l_1,l_2)=(r^2-1)\left(\frac{l_1}{r+l_1}+\frac{l_2}{r+l_2}-\frac{l_1+l_2}{r+l_1+l_2} \right).
\end{equation}
In the case $l_1=1$ we have the conformal field theory with the central charge
\begin{equation}\label{W_r_cent_ch}
c(r,1,l_2)=(r-1)\frac{l_2(2r+l_2+1)}{(r+l_2)(r+l_2+1)},
\end{equation}
which has the $W_r$-symmetry \cite{Fateev:1987zh}. Then, if $l_2$ is a positive integer, the central charge is given by the same formula (\ref{W_r_cent_ch}) and coset describes the Minimal Model with the $W_r$-symmetry.

Let us now consider the case of rank $r=2$, which is studied in the present paper. The coset (\ref{cft_coset}) takes the form
\begin{equation}\label{cft_coset_r2}
\frac{\widehat{\mathfrak{sl}}(2)_{l_1} \times\widehat{\mathfrak{sl}}(2)_{l_2}}{\widehat{\mathfrak{sl}}(2)_{l_1+l_2}}.
\end{equation}
In the case of arbitrary complex $l_1$ and $l_2$ the coset (\ref{cft_coset_r2}) describes the conformal field theory with the central charge
\begin{equation}
c(2,l_1,l_2)=3\left(\frac{l_1}{l_1+2}+\frac{l_2}{l_2+2}-\frac{l_1+l_2}{l_1+l_2+2} \right).
\end{equation}
In the case $l_1=1$ we have the conformal field theory with the central charge
\begin{equation}\label{Liouv_cent_ch}
c(2,1,l_2)=\frac{l_2(l_2+5)}{(l_2+2)(l_2+3)}=1-\frac{6}{(l_2+2)(l_2+3)},
\end{equation}
which has the Virasoro symmetry. Then, if $l_2$ is a positive integer, the central charge is given by the same formula (\ref{Liouv_cent_ch}) and the coset describes the Minimal Model $\mathcal{M}(l_2+1/l_2+2)$, as it was shown in \cite{Goddard:1984vk,Goddard:1986ee}.

\bibliographystyle{MyStyle}
\bibliography{MyBib}

\end{document}